\documentclass[12pt,preprint]{aastex}

\usepackage{lineno}

\usepackage{graphicx}%
\usepackage{multirow}%
\usepackage{amsmath,amssymb,amsfonts}%
\usepackage{amsthm}%
\usepackage{mathrsfs}%
\usepackage[title]{appendix}%
\usepackage{xcolor}%
\usepackage{textcomp}%
\usepackage{manyfoot}%
\usepackage{booktabs}%
\usepackage{algorithm}%
\usepackage{algorithmicx}%
\usepackage{algpseudocode}%
\usepackage{listings}%

\newcommand{\be}{\begin{equation}}
\newcommand{\ee}{\end{equation}}
\newcommand{\nn}{\mbox{} \nonumber \\ \mbox{} }
\newcommand{\ba}{\begin{eqnarray}}
\newcommand{\ea}{\end{eqnarray}}
\newcommand{\om}{\omega}
\newcommand{\Alfven}{ Alfv\'{e}n }

\newcommand{\E}{{\bf E}}
\newcommand{\B}{{\bf B}}

\renewcommand{\v}{{\bf v}}

\newcommand{\Bf}{{magnetic field}}
\newcommand{\Bfs}{{magnetic fields}}
\newcommand{\NS}{neutron star}
\newcommand{\ms}{magnetosphere}
\newcommand{\NSs}{{neutron stars}}

\newcommand{\Ef}{{electric  field}}

\newcommand{\EM}{electromagnetic}
\newcommand{\BH}{{black hole}}

\newcommand{\Sch}{Schwarzschild}
\newcommand{\mss}{magnetospheres}

\newcommand\eg{\textit{e.g.}}

\newcommand\lo{\mathrel{\raise.3ex\hbox{$<$}\mkern-14mu\lower0.6ex\hbox{$\sim$}}}
\newcommand\go{\mathrel{\raise.3ex\hbox{$>$}\mkern-14mu\lower0.6ex\hbox{$\sim$}}}

\theoremstyle{thmstyleone}%
\theoremstyle{thmstyletwo}%

\theoremstyle{thmstylethree}%

\begin{document}
 
\title{Alfven-winged pulsar}

\author{Maxim Lyutikov\\
Department of Physics and Astronomy, Purdue University, \\
 525 Northwestern Avenue,
West Lafayette, IN
47907-2036 }


\begin{abstract}
Detecting possible electromagnetic precursors to the gravitational signal from merging compact objects is challenging, but it can reveal intricate physical properties of the merging stars through  their gravitational and electromagnetic interactions. We demonstrate, using 3D Particle-In-Cell simulations, that a neutron star moving through the magnetosphere of a merging companion generates a complicated system of dissipative currents, a relativistic analogue of planetary Alfven wings. Generated electric currents carry a large fraction of the electromagnetic power intersected by the neutron star. These currents may lead to the generation of beamed, pulsar-like coherent radio and high-energy emission. Orbital modulation will produce a nearly periodic signal, an  {\it  Alfven-winged pulsar}. 
\end{abstract}

\maketitle

\section{Introduction}

The detection of binary \NS\  (BNS) merge by the  LIGO and Virgo detectors \citep[dubbed GW170817][]{2017PhRvL.119p1101A}, was a triumphant confirmation of the   long-predicted event.
It is highly desirable to detect  possible electromagnetic precursor emission   to the main event associated with the merger of  binary neutron stars BNSs. Detecting precursor emission is challenging, but possible \citep{2024arXiv240216504L}.

Compacts' mergers  (double {\NS}s, DNS, \NS-\BH, and double {\BH}s) are expected to occur in old systems, with not much surrounding material. 
Thus,  little  pre-merger accretion power  is expected \citep{2018PhRvD..98h1501F,2020PhR...886....1N}.

In the case of merging \NSs, it is expected that merger times, at least millions of years after the formation of the second \NS\ are sufficiently long so that \EM\ power generated by each \NS\ independently is not large (\NSs\ with large magnetic fields would spin down to slow rotation rates, while millisecond pulsars with small(ish) magnetic fields are not efficient emitters). Thus, at the time of  merger each \ms\ is likely to be dead on its own - not producing pairs via vacuum break-down due to spins of the \NSs.
As the stars spiral in, the magnetospheres can be revived due to the relative orbital motion of the \mss.

The best hope to generate precursor emission is the \EM\ interaction of two merging \NSs\ via the creation of inductive electric fields by the relative motion of magnetized plasma within the common \ms\ \citep{2001MNRAS.322..695H,2019MNRAS.483.2766L,2020ApJ...893L...6M,2021ApJ...923...13C,2023ApJ...956L..33M}. The resulting emission then might  resemble pulsar magnetospheric emission (still unresolved problem), and/or the  better understood emission from magnetospheric interactions in the case  Jupiter-Solar wind, and/or Galilean satellites (\eg\  Ganymede) - Jovian \ms\ \citep{1969ApJ...156...59G,1998JGR...10319963K}.

In the relativistic case, 
the basic estimate of EM power expected from a linearly moving   unipolar inductor \cite{2002luml.conf..381B} is 
\ba  &&
L_{EM}\sim  \frac{c}{4\pi} (\Delta \Phi)^2
\nn &&
  (\Delta \Phi) \sim  2 \beta  B R_{NS}   \approx 3 \times 10^{17} \, (-t)^{-7/8} {\rm eV} 
  \nn && 
  L_{EM} \sim \frac{ B_{{NS}}^2 G M_{{NS}} R_{{NS}}^8}{c r^7} \approx 3 \times { 10^{43} } \;{(-t)^{-7/4}}\,B_{13}^2\, {\rm \, erg \, s^{-1}} 
\label{L} 
\ea
where 
$ 
 (\Delta \Phi) 
$
 is the EMF drop for a conductor of size $2 R_{NS}$ moving in \Bf\ $B$ with velocity $v = \beta c$;  the time to merger $-t$,
\begin{equation}
- t= \frac{5}{256} \frac{c^5}{G^3} \frac{r^4}{M_1 M_2 (M_1+M_2)},
\label{tm}
 \end{equation}
 is measured in seconds in Eq.  (\ref{L}). The factor $4\pi/c= 377 $ Ohm is sometimes called  the impedance of free space.
 
 For a pulsar spinning at period $P$, the power (\ref{L}) becomes larger than the spin-down power for times before merger
 \be
 -t \leq  \frac{5 c^{43/7} R_{{NS}}^{8/7}}{512 \pi ^{4/7} G^{17/7} M_{NS} ^{17/7} \Omega ^{16/7}} \approx  2000 P^{16/7}\, {\rm sec}
 \ee
 where $P$ is in seconds and we assumed similar surface \Bfs\ for both \NSs. 
 
 Thus, for old \NSs\ with periods in the seconds range, the linear inductive power will be larger than the spindown power $\sim$ tens of minutes before the merger.
 
On the other  hand, the estimates  (\ref{L}), normalized to a relatively large surface magnetic field of $10^{13}$ G, highlight a potential issue: the overall expected power is not particularly large. If emitted isotropically, it will be missed by high-energy all-sky X-ray monitors (if coming form $\geq 10$ Mpc distance).

We may still detect precursor emission. First, if some fraction  $\eta_R = 10^{-3} \eta_{R,-3} $,  of power (\ref{L}) is emitted in radio  the expected observed flux may be in the Jansky range  \citep{2001MNRAS.322..695H,2019MNRAS.483.2766L,2023MNRAS.519.3923C}
\be
F_{\nu, peak} = 0.5 \, {\rm Jy}  \eta_{R,-3}  \nu_{9} ^{-1} d_{200}^{-2}
\ee
where $\nu = 10^ 9 \nu_{9}  $ Hz is the observed frequency and $d_{200} =d/(200) \, {\rm Mpc}$ is distance to the source. 
Second, if the power  (\ref{L}) is beamed, the  observed peak flux would  higher. (Radio detection will also need modulation).
As we demonstrate in the paper, the power from the merging \NSs\ is indeed expected to be beamed  due to formation of relativistic \Alfven\ wings.


\section{\Alfven wings -  from planets to merging compact objects.}

The interaction of celestial bodies with magnetized plasma is a classical problem in space physics: Earth \ms-Solar wind is the most studied case. The interaction of merging \NSs\ will occur in  a new, so-far unexplored regime of highly magnetized plasma, and (mildly) relativistically moving companions. 

In space plasma environments,  the interaction of   bodies with Solar wind or planetary \ms\ displays markedly different behavior depending on (i) intrinsic \Bf, (ii)   conductivity of a moving body (planets vary greatly in their electrical conductivity); (iii) flow regime. 

One of the key structures are \Alfven wings,  observed when a conducting obstacle is embedded in a sub-Alfvenic flow. (Occasionally,  during rare periods of extremely low-density solar wind, the Earth's magnetosphere has been observed forming \Alfven wing \cite{2025GeoRL..5211931G}.) Similar processes occur at the Moon, although low   conductivity and lack of an intrinsic magnetic field leads to plasma wake formation rather than stable wings \cite{2012JGRA..117.9217C, 2016JGRA..12110698Z}. The most direct analogues for strongly conducting bodies are Jupiter's moons. Io's volcanically sustained ionosphere and Ganymede's intrinsic magnetic field each generate coherent \Alfven wings and associated field-aligned current systems \cite{2017P&SS..137...40V, 1999JGR...10428671N, 2002JGRA..107.1490K}, which are observable through auroral and radio signatures \cite{2014A&A...569A..86M, 2024EPSC...17..726C}. Particularly relevant is Io-modulated  Jovian decametric radiation  \citep{1978JGR....83.2617S}.

Extensions of \Alfven\ wing theory to relativistic flows have been explored in the context of pulsar winds and orbiting bodies by \cite{2011A&A...532A..21M}, demonstrating that strong currents can be driven even when the body is immersed in a relativistic plasma flow. These examples clearly demonstrate that conducting obstacles in magnetized plasma flows produce electromagnetic structures that efficiently channel energy and can be detected observationally.

Neutron stars represent the extreme limit of this physics. Their matter composition renders them effectively perfect conductors, enforcing a boundary condition that expels magnetic flux from the stellar interior. As a result, magnetic field lines drape tightly around the surface, forming thin current layers whose structure depends on the flow velocity regime. In non-relativistic flows, these layers primarily amplify magnetic fields through compression. In relativistic flows, however, the interaction becomes qualitatively different: electric fields become independent dynamical variables, and the draping layer can develop dissipative regions where electromagnetic energy is converted to particle acceleration and radiation.

Magnetic interaction of merging \NSs\ is expected to proceed in an unusual (by space physics) regime: (i) plasma is expected to be highly magnetized, with $\sigma = (\om_B/\om_p)^2  \geq 1$ (and,  correspondingly  plasma beta $\ll 1$; (ii) interaction velocities may be mildly relativistic $\sim c$, as near the merger the Keplerian velocity is close to the speed of light.  Thus, the interaction may be relativistic but sub-Alfvenic. This unusual regime is the goal of the present study.


In the present  work, we   examine the electromagnetic interaction between a perfectly conducting, unmagnetized neutron star and the magnetized plasma of its companion. The core mechanism governing this interaction is electromagnetic draping, in which ambient magnetic field lines accumulate and compress near the moving conductor. This creates thin boundary layers where strong currents flow and electromagnetic energy can be dissipated \cite{2004AIPC..719..381C,2006MNRAS.373...73L,2008ApJ...677..993D}. 
PIC simulations are needed to resolve the dissipative structures within the flow.

\section{Basic results: formation  of Alfven wing structures}

For relativistic flows, the (trans)-Alfvenic regime is demarcated by the condition on four-velocities, \Alfven\ four-velocity $u_A = \sqrt{\sigma} $ being equal the flow  four-velocity $u$,   $u_A=u$. Parameter sigma, $\sigma = \om_B^2/\om_p^2$,  is plasma magnetization.

For our  basic run, Appendic \ref{EPOCH},  the parameters evaluate to   $\sigma \approx 5$ ($\sigma \approx 10$ for each species separately). $\beta_0 = 0.5$ (so that $\gamma_A = \sqrt{1+\sigma} \approx 2.44 > \gamma = 1.15$. The flow is mildly relativistic, but sub-Alfvenic, $M_A= 0.26$.

\begin{figure}[h!]
 \includegraphics[width=.99\linewidth]{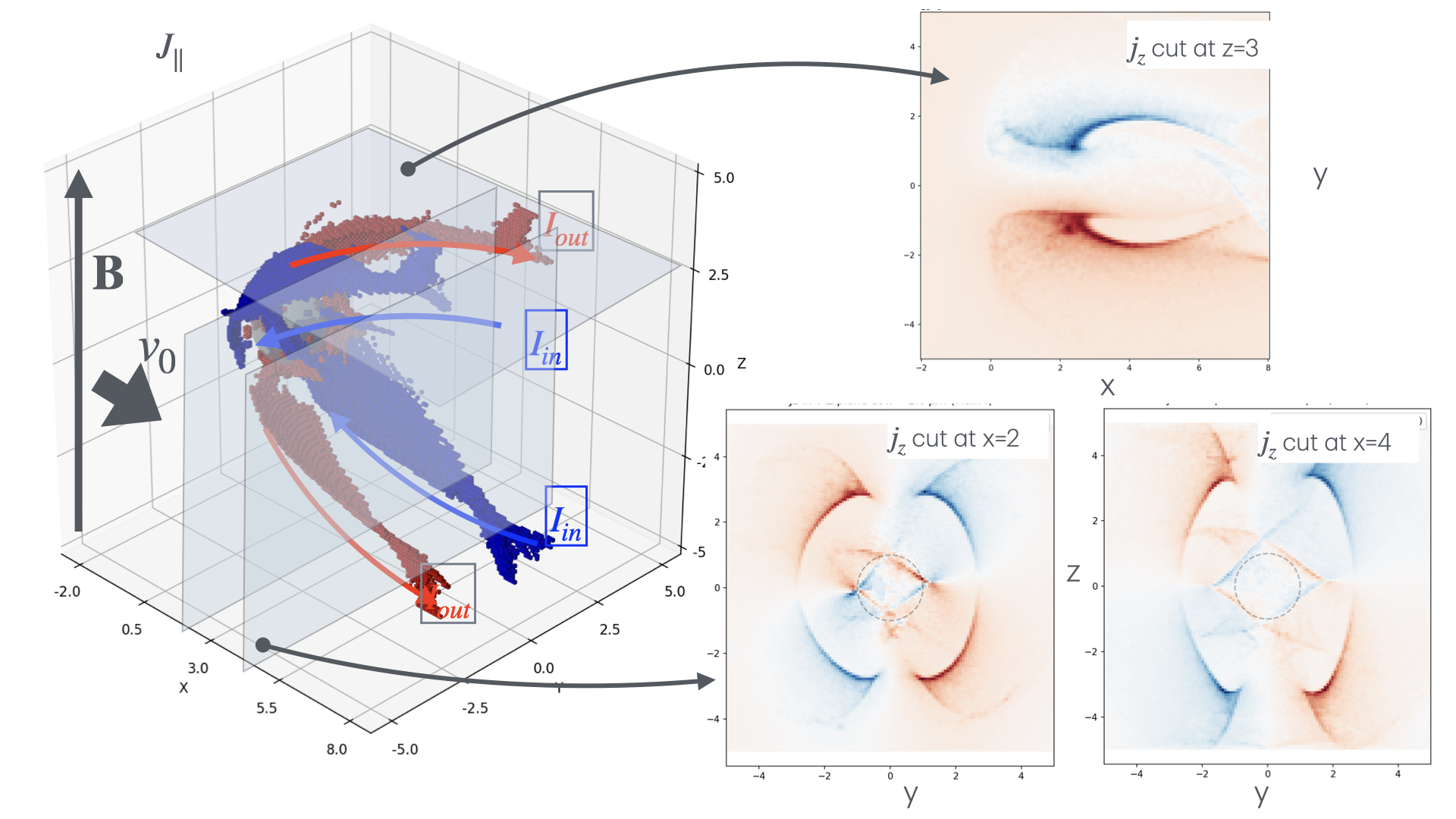} 
  \caption{(Annotated) Isosurfaces of current $j_\parallel = {\bf J} \cdot {\bf B}/|B|$,  and slices  of $j_z$ at fixed $z=3$ (top left)  and fixed $x=2,\,4$.  Red and blue arrows indicate the current flow. The figure demonstrates complicated, large-scale  current/\EM\ structure in the wake.} 
 \label{iso-1}
\end{figure}

In Fig.  \ref{iso-1}  we plot isosurfaces of parallel current ${\bf j}\cdot {\bf B}/|B|$ (at value of 0.1 of the maximal). We clearly observe formation of  \Alfven\ wings: X-type (double-jetted)  structure of the current. Direction of the total  current reverses in each quadrant.

\begin{figure}[h!]
  \includegraphics[width=.99\linewidth]{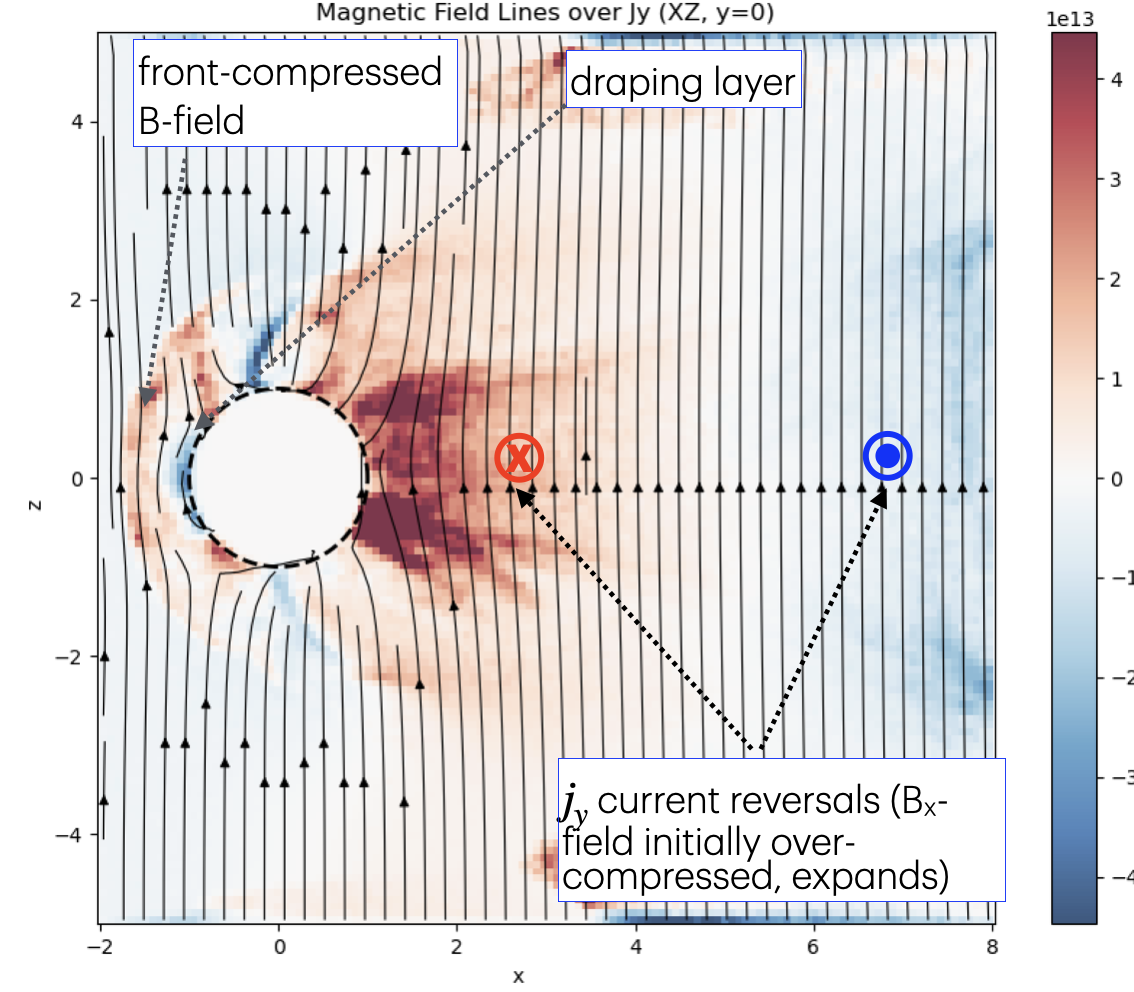}
  \caption{The x-z slice   (y=0) of magnetic field lines and current $j_y$ (out of the board) annotated.}
 \label{jy-B}
\end{figure}

 Fig. \ref{jy-B} show a structure of the \Bf\ and the current $j_y$ (out of the board). It reveals  a highly intricate, complicated interplay between plasma, currents  and \EM\ fields in the  interaction wake. There are reversing currents at the head part, $ -2<x < -1$,  associated with plasma compression (diamagnetic currents, magnetic draping). 
 
  Some of the features we observe are not expected: \eg,   there are polar wings,  tail-ward  current enhancements and large-scale reversals of $j_y$.  (In the tail part there is first high $j_y$ currents out of the board, switching downstream to $j_y$ currents  into the board.

 Most importantly, these figurers demonstrate that the ensuing interaction is non-local, expending to scales much larger than the size of the conducting object.  
Somewhat unexpectedly, there are features appearing at scales $\gg R$ (\eg\ bending of isosurfaces $x\sim 5$). These features are physical.

\begin{figure}[h!]
  \includegraphics[width=.99\linewidth]{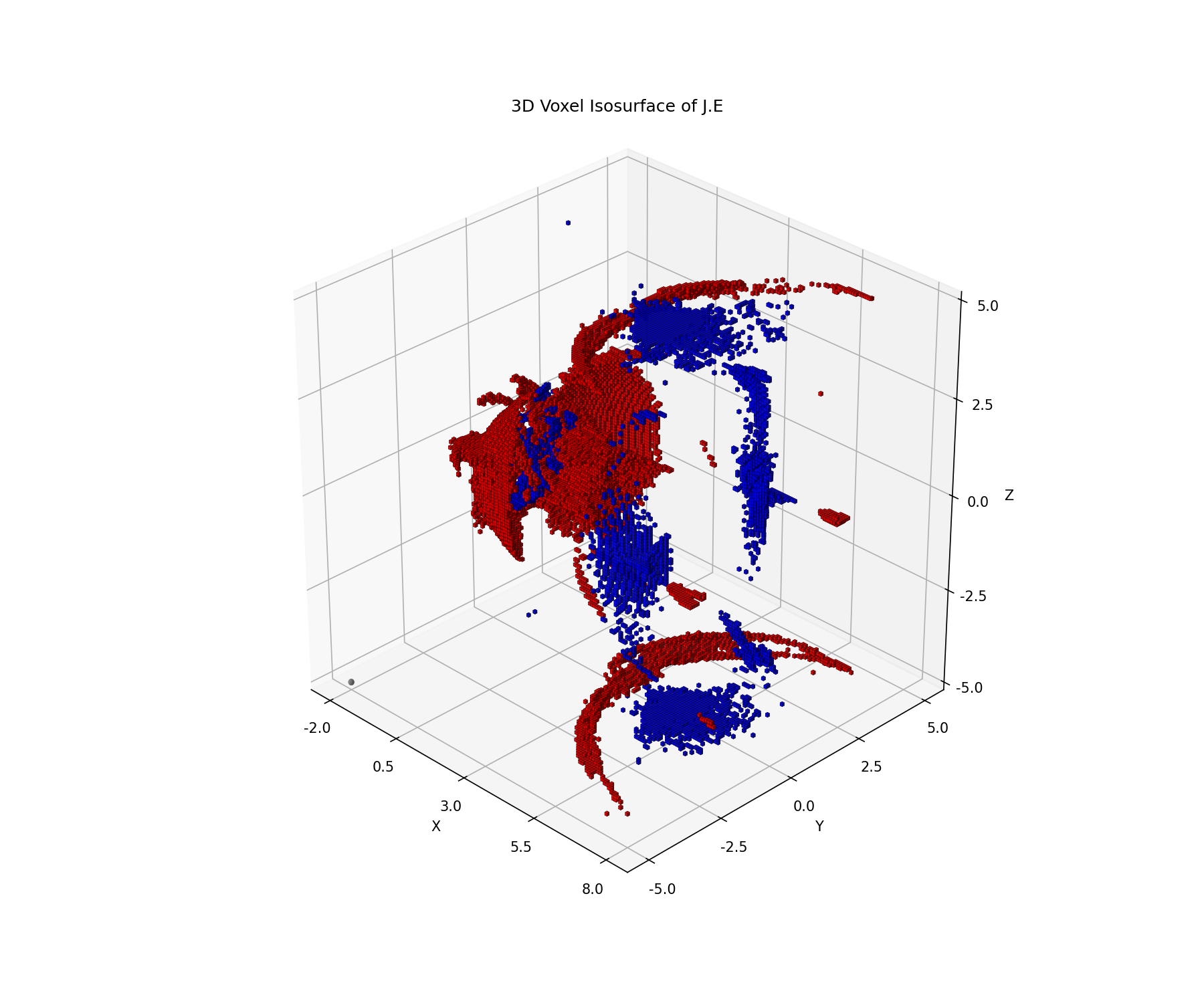}
  \caption{3D rendering of ${\bf J} \cdot {\bf E}$, regions of intensive  plasma-\EM\  energy exchange. }
 \label{JE}
\end{figure}

Fig. \ref{JE} illustrates plasma-\EM\ energy exchange by plotting  ${\bf J} \cdot {\bf E}$. One observes large-scale interaction structures. 

\begin{figure}[h!]
 \includegraphics[width=.49\linewidth]{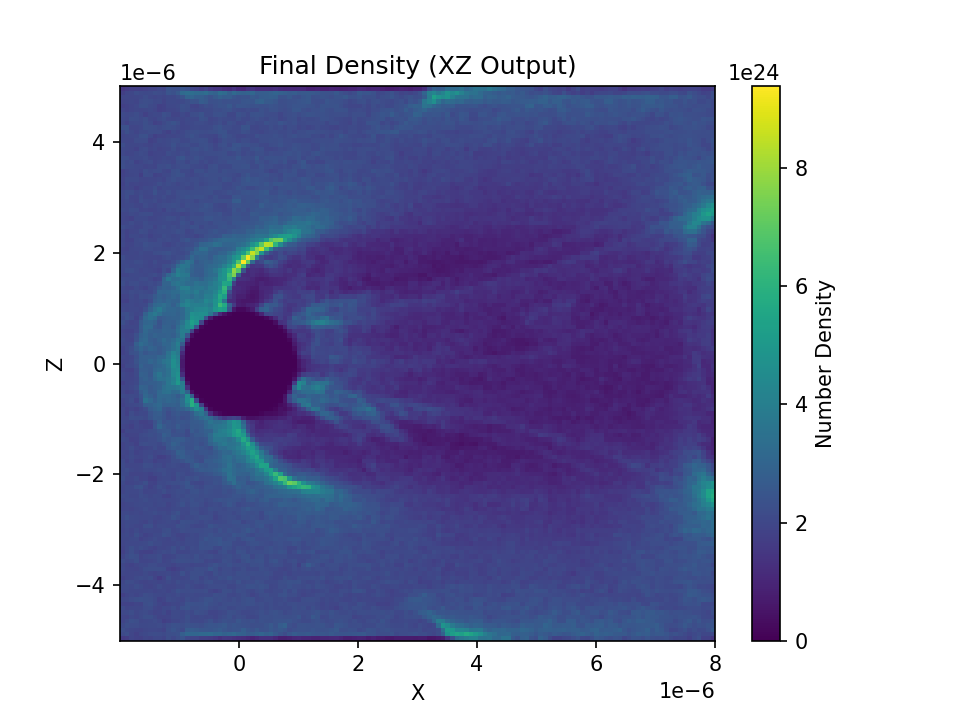} 
  \includegraphics[width=.49\linewidth]{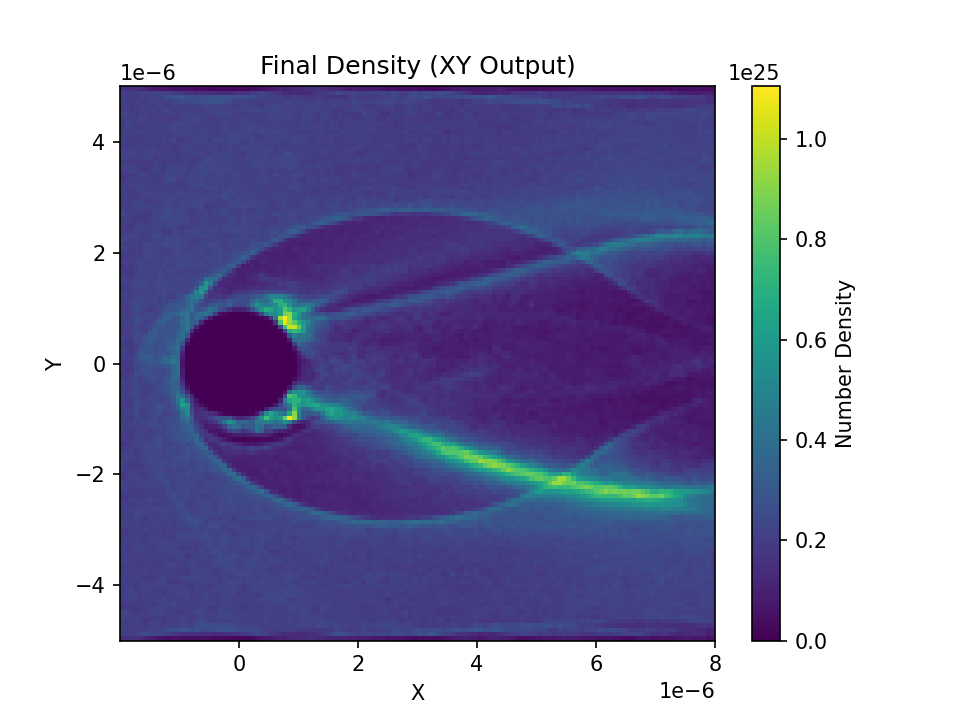} \\
     \includegraphics[width=.49\linewidth]{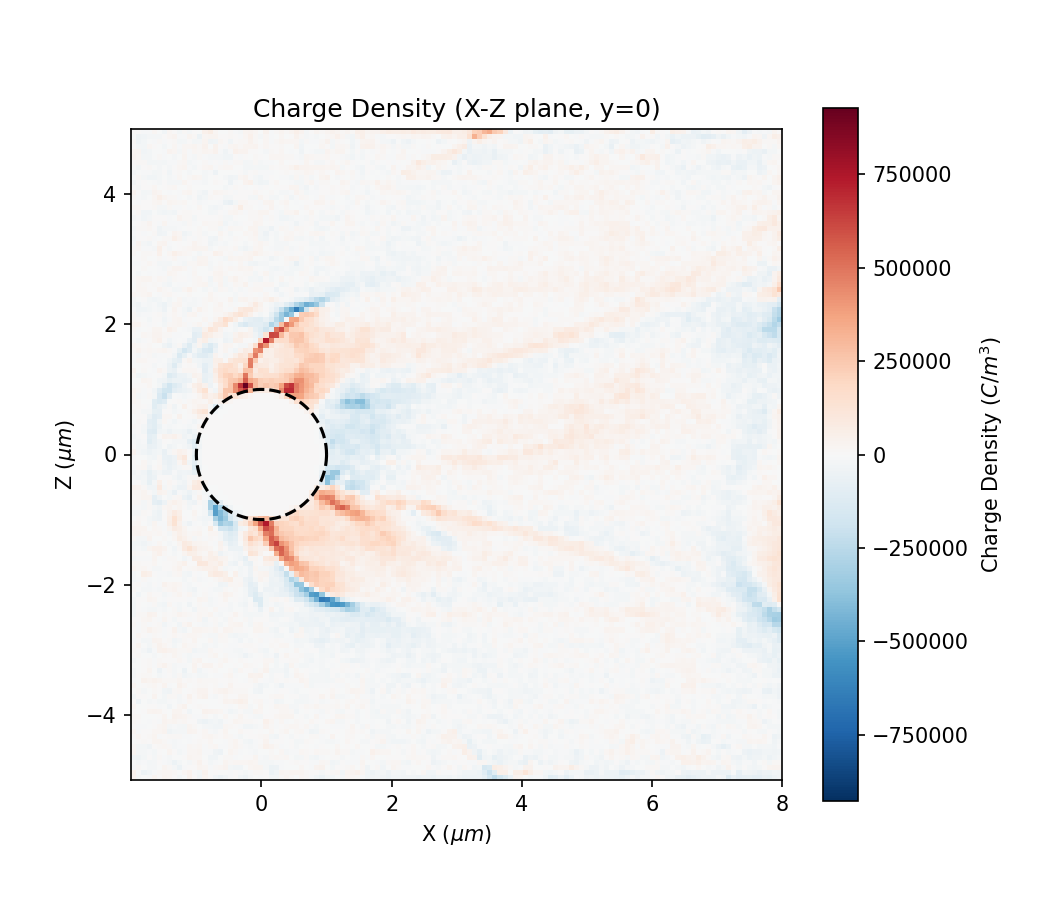} 
   \includegraphics[width=.49\linewidth]{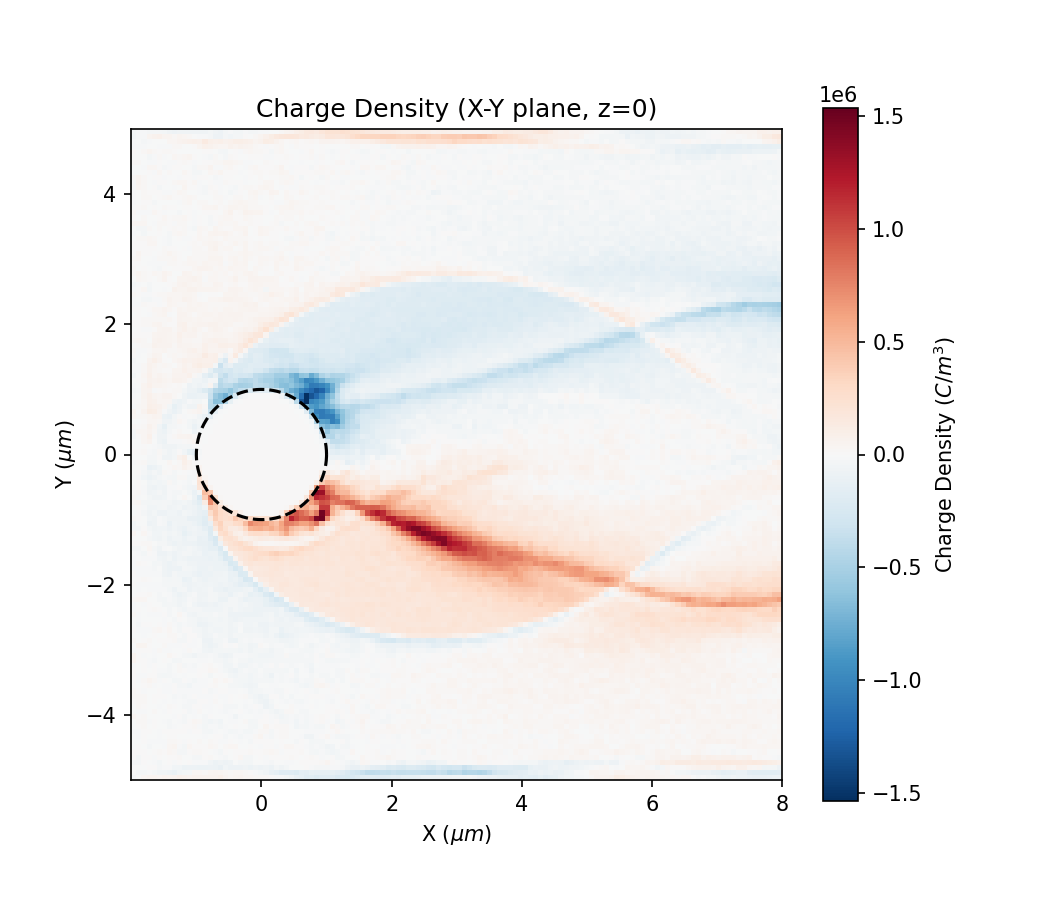} 
      \includegraphics[width=.45\linewidth]{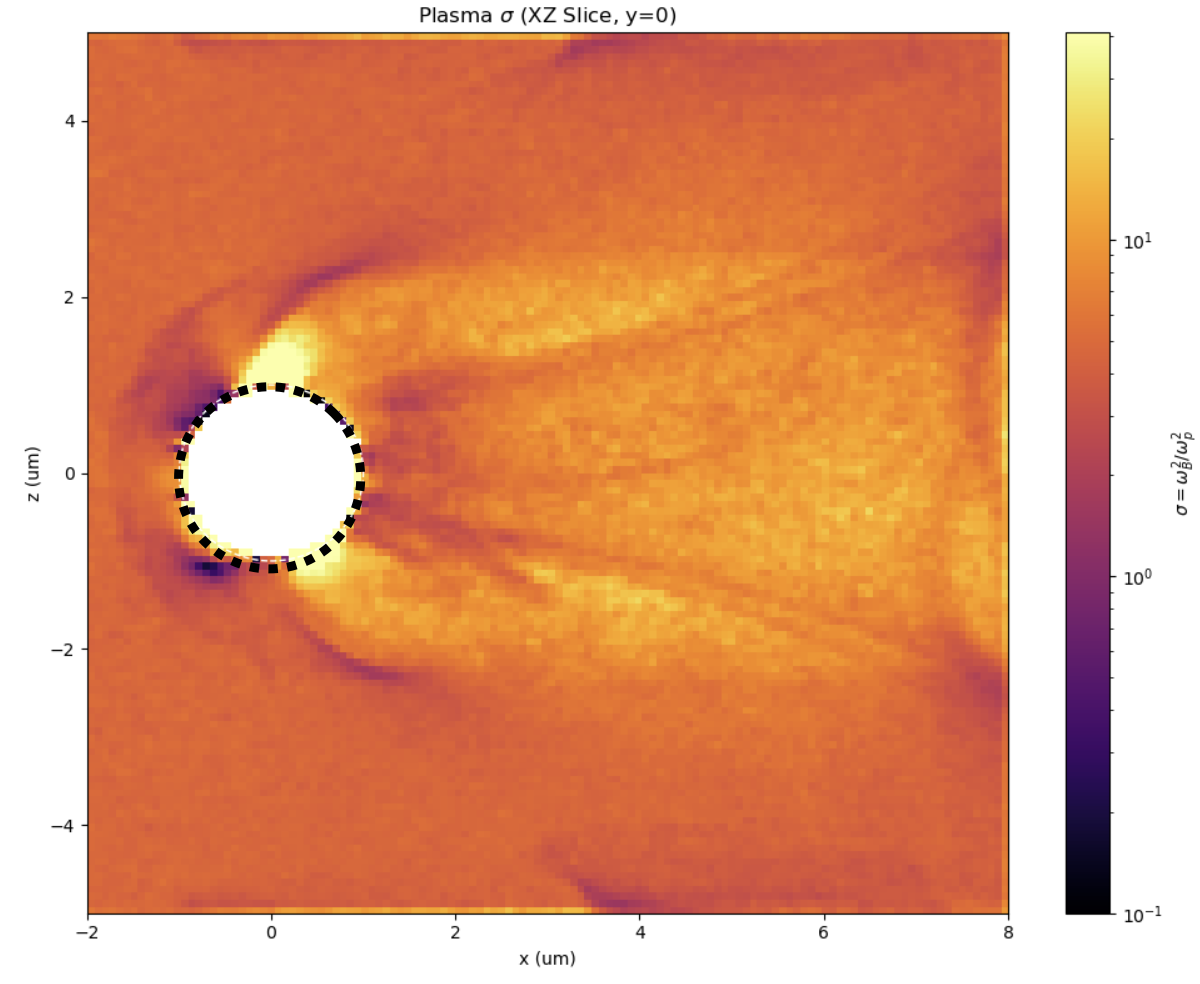} 
   \includegraphics[width=.45\linewidth]{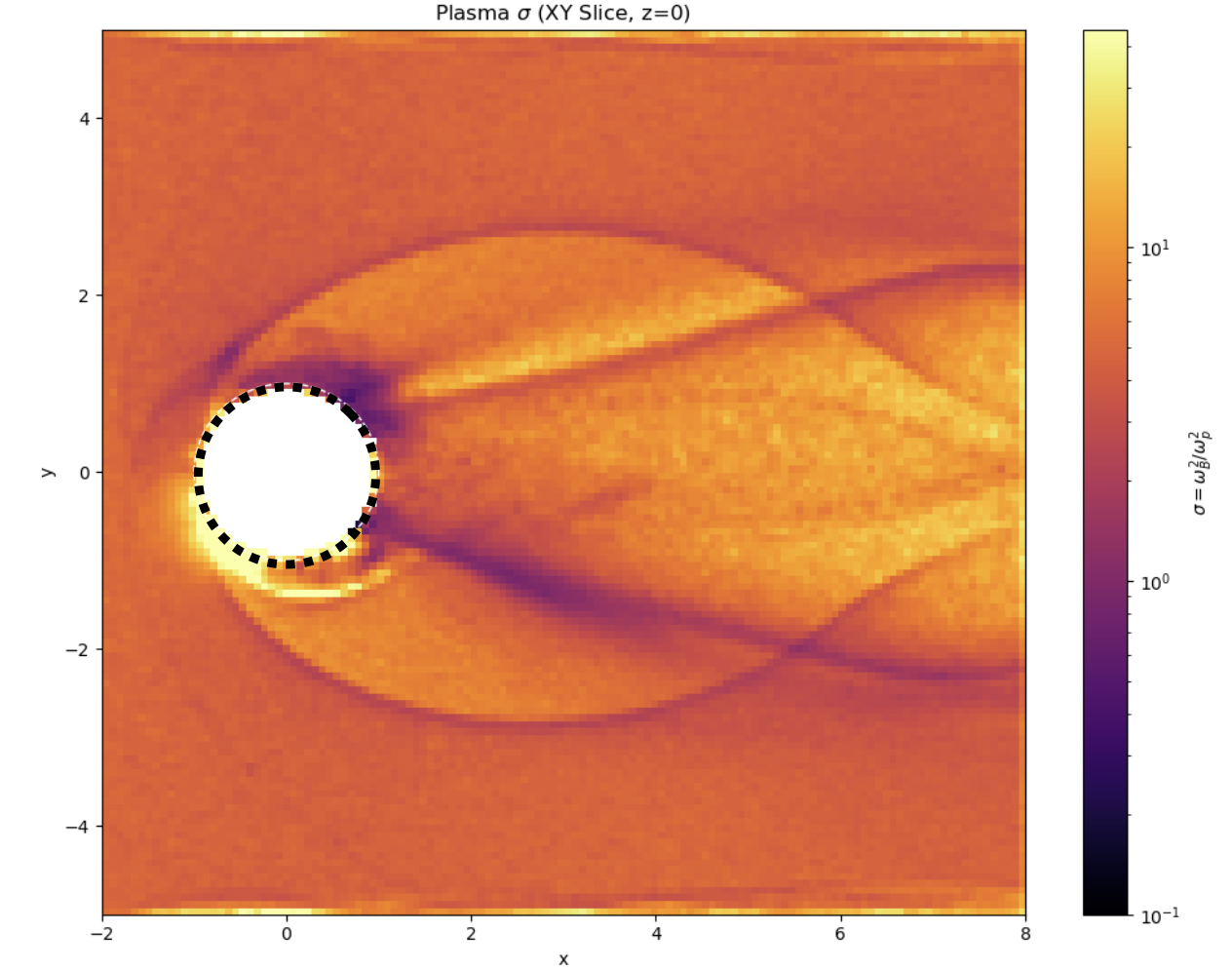} 
  \caption{X-Z (left) and X-Y (right)   slices;  Tot row: density, middle row: charge density, bottom row plasma sigma $\sigma = \om_B^2 /\om_p^2$. Slight asymmetry in the  X-Y plane is likely due to mild left-right instability of the flow.  Non-negligible charge density within the \Alfven\ wings is  specific to the relativistic regime.}. 
 \label{slices1}
\end{figure}

Finally, in Fig. \ref{slices1} we plot density and charge density slices (vertical and horizontal). Both are highly revealing: they highlight eh large-scale structure of relativist charges and currents. Non-negligible charge density is a specific feature of relativistic \Alfven\ wings.

We have verified that  the flow vorticity is concentrated near the surface of the NS, the bulk flow is nearly potential.
(In the density plots, the  lines at the back of the sphere  that look like vortex shedding, they  are not.)

In the highly relativistic regime, $\sigma \gg 1$, it is expected that the power in the Alfven wings $L_w$,  and  the corresponding current $I_w$  scale in proportion to the inductive potential, 
\be
L_w \propto  I_w^2  \propto (B_0 v_0)^2
\ee

For our basic run, we have measured $I_w$ as  total current) in the slice $z=\pm3$. The average value is zero.  Integration over  $\pm y$  half-planes are identical, different just by the   sigs. 
We have verified that $I_w$ currents in the four quadrants  ($z=3, y\pm 0$, ($z=-3, y\pm 0$) are  equal in absolute magnitudes. 

We have measured the total current  $I_w$ at  $z= \pm 3 $ and $ y>< 0$.  The  efficiency of  \EM\ jets is approximately (see Appendix \ref{EPOCH})
 \be
\eta = \left(  \frac{ 2 \times I_w}{ I_0}  \right) ^2 =0.22
\ee
Thus, a large fraction of the \EM\ Poynting flux impinging on the \NS\ is converted to the currents supporting \Alfven\ wings.

\section{Discussion}

We address a classical problem in space physics -- a conducting ball moving through magnetized plasma -- in an astrophysically  relevant relativistic regime. Guided by the theoretical insight \citep{2023PhRvE.107b5205L} and results of MHD results \cite{Praveen} we perform 3D PIC simulations.   The seemly simple  system turns out to be  rich in intricate details: complicated 3D current system is formed, \Bf-aligned currents are generated.

Analytics  \citep{2023PhRvE.107b5205L}  predicts formation of dissipative regions. MHD simulations \cite[in the ideal limit -  no explicit resistivity][]{Praveen}  show  the formation of \Alfven\ wings. The present  PICs  simulations resolve kinetic details of \Alfven\ wings formation, \eg\ non-zero values of ${\bf E} \cdot  {\bf j} $, which is  missed from  ideal MHS codes. The similarity of the results between  PIC and MHD  simulations  then validates the MHD simulations: ideal MHD code correctly resolves the non-ideal effects in the draping layer.

We have also verified that motion along \Bf\ does not produce \Alfven\ wings structure - only mildly dissipative wake (for our basic parameters) is observed. We expect that for super-Alfvenic runs the cases of parallel and perpendicular \Bfs\ will be similar. \citep[In passing we note that motion along \Bf\ is not necessarily  electromagnetically dead, \eg][]{2025PhRvD.111j3010K}

 An obvious next step is the  simulations of the interaction of a \Sch\  BH moving through  magnetized plasma. Theory  \cite{2011PhRvD..83f4001L} and early MHD simulations  \citep{2010Sci...329..927P} predict formation of \Alfven\ wings.  A glimpse at the expected results can be obtained in flat space 
by changing the boundary conditions on the surface of  an object from {\it conducting} to {\it open} (and removing internal  particles  and fields),   Appendix \ref{MimickingBH}. 
This flat-space run mimics the interaction of a \BH\ with magnetized plasma.  Somewhat surprisingly both NS and BH cases look qualitatively similar. This exercise bears important lesson: formation of \Alfven\ wings is generally independent of the details of \Bf\ draping.

We would like to thank Anatoly Spitkovsky for numerous discussions and code advice. 
 \bibliographystyle{apj} 
 \bibliography{/Users/lyutikov/Library/CloudStorage/Dropbox/Research/BibTex,/Users/lyutikov/Library/CloudStorage/Dropbox/Research/BibTexShort.bib,//Users/lyutikov/Library/CloudStorage/Dropbox/Research/NASA_FRB.bib,/Users/lyutikov/Library/CloudStorage/Dropbox/Research/references} 

\begin{thebibliography}{31}
\expandafter\ifx\csname natexlab\endcsname\relax\def\natexlab#1{#1}\fi
\expandafter\ifx\csname bibnamefont\endcsname\relax
  \def\bibnamefont#1{#1}\fi
\expandafter\ifx\csname bibfnamefont\endcsname\relax
  \def\bibfnamefont#1{#1}\fi
\expandafter\ifx\csname citenamefont\endcsname\relax
  \def\citenamefont#1{#1}\fi
\expandafter\ifx\csname url\endcsname\relax
  \def\url#1{\texttt{#1}}\fi
\expandafter\ifx\csname urlprefix\endcsname\relax\def\urlprefix{URL }\fi
\providecommand{\bibinfo}[2]{#2}
\providecommand{\eprint}[2][]{\url{#2}}

\bibitem[{\citenamefont{{Abbott} et~al.}(2017)\citenamefont{{Abbott}, {Abbott},
  {Abbott}, {Acernese}, {Ackley}, {Adams}, {Adams}, {Addesso}, {Adhikari},
  {Adya} et~al.}}]{2017PhRvL.119p1101A}
\bibinfo{author}{\bibfnamefont{B.~P.} \bibnamefont{{Abbott}}},
  \bibinfo{author}{\bibfnamefont{R.}~\bibnamefont{{Abbott}}},
  \bibinfo{author}{\bibfnamefont{T.~D.} \bibnamefont{{Abbott}}},
  \bibinfo{author}{\bibfnamefont{F.}~\bibnamefont{{Acernese}}},
  \bibinfo{author}{\bibfnamefont{K.}~\bibnamefont{{Ackley}}},
  \bibinfo{author}{\bibfnamefont{C.}~\bibnamefont{{Adams}}},
  \bibinfo{author}{\bibfnamefont{T.}~\bibnamefont{{Adams}}},
  \bibinfo{author}{\bibfnamefont{P.}~\bibnamefont{{Addesso}}},
  \bibinfo{author}{\bibfnamefont{R.~X.} \bibnamefont{{Adhikari}}},
  \bibinfo{author}{\bibfnamefont{V.~B.} \bibnamefont{{Adya}}},
  \bibnamefont{et~al.}, \bibinfo{journal}{Physical Review Letters}
  \textbf{\bibinfo{volume}{119}}, \bibinfo{eid}{161101} (\bibinfo{year}{2017}),
  \eprint{1710.05832}.

\bibitem[{\citenamefont{{Lyutikov}}(2024)}]{2024arXiv240216504L}
\bibinfo{author}{\bibfnamefont{M.}~\bibnamefont{{Lyutikov}}},
  \bibinfo{journal}{arXiv e-prints} \bibinfo{eid}{arXiv:2402.16504}
  (\bibinfo{year}{2024}), \eprint{2402.16504}.

\bibitem[{\citenamefont{{Foucart} et~al.}(2018)\citenamefont{{Foucart},
  {Hinderer}, and {Nissanke}}}]{2018PhRvD..98h1501F}
\bibinfo{author}{\bibfnamefont{F.}~\bibnamefont{{Foucart}}},
  \bibinfo{author}{\bibfnamefont{T.}~\bibnamefont{{Hinderer}}},
  \bibnamefont{and}
  \bibinfo{author}{\bibfnamefont{S.}~\bibnamefont{{Nissanke}}},
  \bibinfo{journal}{\prd} \textbf{\bibinfo{volume}{98}}, \bibinfo{eid}{081501}
  (\bibinfo{year}{2018}), \eprint{1807.00011}.

\bibitem[{\citenamefont{{Nakar}}(2020)}]{2020PhR...886....1N}
\bibinfo{author}{\bibfnamefont{E.}~\bibnamefont{{Nakar}}},
  \bibinfo{journal}{\physrep} \textbf{\bibinfo{volume}{886}},
  \bibinfo{pages}{1} (\bibinfo{year}{2020}), \eprint{1912.05659}.

\bibitem[{\citenamefont{{Hansen} and {Lyutikov}}(2001)}]{2001MNRAS.322..695H}
\bibinfo{author}{\bibfnamefont{B.~M.~S.} \bibnamefont{{Hansen}}}
  \bibnamefont{and}
  \bibinfo{author}{\bibfnamefont{M.}~\bibnamefont{{Lyutikov}}},
  \bibinfo{journal}{\mnras} \textbf{\bibinfo{volume}{322}},
  \bibinfo{pages}{695} (\bibinfo{year}{2001}), \eprint{astro-ph/0003218}.

\bibitem[{\citenamefont{{Lyutikov}}(2019)}]{2019MNRAS.483.2766L}
\bibinfo{author}{\bibfnamefont{M.}~\bibnamefont{{Lyutikov}}},
  \bibinfo{journal}{\mnras} \textbf{\bibinfo{volume}{483}},
  \bibinfo{pages}{2766} (\bibinfo{year}{2019}).

\bibitem[{\citenamefont{{Most} and {Philippov}}(2020)}]{2020ApJ...893L...6M}
\bibinfo{author}{\bibfnamefont{E.~R.} \bibnamefont{{Most}}} \bibnamefont{and}
  \bibinfo{author}{\bibfnamefont{A.~A.} \bibnamefont{{Philippov}}},
  \bibinfo{journal}{\apjl} \textbf{\bibinfo{volume}{893}}, \bibinfo{eid}{L6}
  (\bibinfo{year}{2020}), \eprint{2001.06037}.

\bibitem[{\citenamefont{{Cherkis} and {Lyutikov}}(2021)}]{2021ApJ...923...13C}
\bibinfo{author}{\bibfnamefont{S.~A.} \bibnamefont{{Cherkis}}}
  \bibnamefont{and}
  \bibinfo{author}{\bibfnamefont{M.}~\bibnamefont{{Lyutikov}}},
  \bibinfo{journal}{\apj} \textbf{\bibinfo{volume}{923}}, \bibinfo{eid}{13}
  (\bibinfo{year}{2021}), \eprint{2107.09702}.

\bibitem[{\citenamefont{{Goldreich} and
  {Lynden-Bell}}(1969)}]{1969ApJ...156...59G}
\bibinfo{author}{\bibfnamefont{P.}~\bibnamefont{{Goldreich}}} \bibnamefont{and}
  \bibinfo{author}{\bibfnamefont{D.}~\bibnamefont{{Lynden-Bell}}},
  \bibinfo{journal}{\apj} \textbf{\bibinfo{volume}{156}}, \bibinfo{pages}{59}
  (\bibinfo{year}{1969}).

\bibitem[{\citenamefont{{Kivelson} et~al.}(1998)\citenamefont{{Kivelson},
  {Warnecke}, {Bennett}, {Joy}, {Khurana}, {Linker}, {Russell}, {Walker}, and
  {Polanskey}}}]{1998JGR...10319963K}
\bibinfo{author}{\bibfnamefont{M.~G.} \bibnamefont{{Kivelson}}},
  \bibinfo{author}{\bibfnamefont{J.}~\bibnamefont{{Warnecke}}},
  \bibinfo{author}{\bibfnamefont{L.}~\bibnamefont{{Bennett}}},
  \bibinfo{author}{\bibfnamefont{S.}~\bibnamefont{{Joy}}},
  \bibinfo{author}{\bibfnamefont{K.~K.} \bibnamefont{{Khurana}}},
  \bibinfo{author}{\bibfnamefont{J.~A.} \bibnamefont{{Linker}}},
  \bibinfo{author}{\bibfnamefont{C.~T.} \bibnamefont{{Russell}}},
  \bibinfo{author}{\bibfnamefont{R.~J.} \bibnamefont{{Walker}}},
  \bibnamefont{and}
  \bibinfo{author}{\bibfnamefont{C.}~\bibnamefont{{Polanskey}}},
  \bibinfo{journal}{\jgr} \textbf{\bibinfo{volume}{103}},
  \bibinfo{pages}{19963} (\bibinfo{year}{1998}).

\bibitem[{\citenamefont{{Blandford}}(2002)}]{2002luml.conf..381B}
\bibinfo{author}{\bibfnamefont{R.~D.} \bibnamefont{{Blandford}}}, in
  \emph{\bibinfo{booktitle}{Lighthouses of the Universe: The Most Luminous
  Celestial Objects and Their Use for Cosmology}}, edited by
  \bibinfo{editor}{\bibfnamefont{M.}~\bibnamefont{{Gilfanov}}},
  \bibinfo{editor}{\bibfnamefont{R.}~\bibnamefont{{Sunyeav}}},
  \bibnamefont{and}
  \bibinfo{editor}{\bibfnamefont{E.}~\bibnamefont{{Churazov}}}
  (\bibinfo{year}{2002}), p. \bibinfo{pages}{381}, \eprint{astro-ph/0202265}.

\bibitem[{\citenamefont{{Cooper} et~al.}(2023)\citenamefont{{Cooper}, {Gupta},
  {Wadiasingh}, {Wijers}, {Boersma}, {Andreoni}, {Rowlinson}, and
  {Gourdji}}}]{2023MNRAS.519.3923C}
\bibinfo{author}{\bibfnamefont{A.~J.} \bibnamefont{{Cooper}}},
  \bibinfo{author}{\bibfnamefont{O.}~\bibnamefont{{Gupta}}},
  \bibinfo{author}{\bibfnamefont{Z.}~\bibnamefont{{Wadiasingh}}},
  \bibinfo{author}{\bibfnamefont{R.~A.~M.~J.} \bibnamefont{{Wijers}}},
  \bibinfo{author}{\bibfnamefont{O.~M.} \bibnamefont{{Boersma}}},
  \bibinfo{author}{\bibfnamefont{I.}~\bibnamefont{{Andreoni}}},
  \bibinfo{author}{\bibfnamefont{A.}~\bibnamefont{{Rowlinson}}},
  \bibnamefont{and}
  \bibinfo{author}{\bibfnamefont{K.}~\bibnamefont{{Gourdji}}},
  \bibinfo{journal}{\mnras} \textbf{\bibinfo{volume}{519}},
  \bibinfo{pages}{3923} (\bibinfo{year}{2023}), \eprint{2210.17205}.

\bibitem[{\citenamefont{{Gurram} et~al.}(2025)\citenamefont{{Gurram},
  {Shuster}, {Chen}, {Hasegawa}, {Denton}, {Burkholder}, {Beedle}, {Gershman},
  and {Burch}}}]{2025GeoRL..5211931G}
\bibinfo{author}{\bibfnamefont{H.}~\bibnamefont{{Gurram}}},
  \bibinfo{author}{\bibfnamefont{J.~R.} \bibnamefont{{Shuster}}},
  \bibinfo{author}{\bibfnamefont{L.-J.} \bibnamefont{{Chen}}},
  \bibinfo{author}{\bibfnamefont{H.}~\bibnamefont{{Hasegawa}}},
  \bibinfo{author}{\bibfnamefont{R.~E.} \bibnamefont{{Denton}}},
  \bibinfo{author}{\bibfnamefont{B.~L.} \bibnamefont{{Burkholder}}},
  \bibinfo{author}{\bibfnamefont{J.}~\bibnamefont{{Beedle}}},
  \bibinfo{author}{\bibfnamefont{D.~J.} \bibnamefont{{Gershman}}},
  \bibnamefont{and} \bibinfo{author}{\bibfnamefont{J.}~\bibnamefont{{Burch}}},
  \bibinfo{journal}{\grl} \textbf{\bibinfo{volume}{52}},
  \bibinfo{pages}{2024GL111931} (\bibinfo{year}{2025}), \eprint{2409.00247}.

\bibitem[{\citenamefont{{Chan{\'e}} et~al.}(2012)\citenamefont{{Chan{\'e}},
  {Saur}, {Neubauer}, {Raeder}, and {Poedts}}}]{2012JGRA..117.9217C}
\bibinfo{author}{\bibfnamefont{E.}~\bibnamefont{{Chan{\'e}}}},
  \bibinfo{author}{\bibfnamefont{J.}~\bibnamefont{{Saur}}},
  \bibinfo{author}{\bibfnamefont{F.~M.} \bibnamefont{{Neubauer}}},
  \bibinfo{author}{\bibfnamefont{J.}~\bibnamefont{{Raeder}}}, \bibnamefont{and}
  \bibinfo{author}{\bibfnamefont{S.}~\bibnamefont{{Poedts}}},
  \bibinfo{journal}{Journal of Geophysical Research (Space Physics)}
  \textbf{\bibinfo{volume}{117}}, \bibinfo{eid}{A09217} (\bibinfo{year}{2012}).

\bibitem[{\citenamefont{{Zhang} et~al.}(2016)\citenamefont{{Zhang}, {Khurana},
  {Kivelson}, {Fatemi}, {Holmstr{\"o}m}, {Angelopoulos}, {Jia}, {Wan}, {Liu},
  {Chen} et~al.}}]{2016JGRA..12110698Z}
\bibinfo{author}{\bibfnamefont{H.}~\bibnamefont{{Zhang}}},
  \bibinfo{author}{\bibfnamefont{K.~K.} \bibnamefont{{Khurana}}},
  \bibinfo{author}{\bibfnamefont{M.~G.} \bibnamefont{{Kivelson}}},
  \bibinfo{author}{\bibfnamefont{S.}~\bibnamefont{{Fatemi}}},
  \bibinfo{author}{\bibfnamefont{M.}~\bibnamefont{{Holmstr{\"o}m}}},
  \bibinfo{author}{\bibfnamefont{V.}~\bibnamefont{{Angelopoulos}}},
  \bibinfo{author}{\bibfnamefont{Y.~D.} \bibnamefont{{Jia}}},
  \bibinfo{author}{\bibfnamefont{W.~X.} \bibnamefont{{Wan}}},
  \bibinfo{author}{\bibfnamefont{L.~B.} \bibnamefont{{Liu}}},
  \bibinfo{author}{\bibfnamefont{Y.~D.} \bibnamefont{{Chen}}},
  \bibnamefont{et~al.}, \bibinfo{journal}{Journal of Geophysical Research
  (Space Physics)} \textbf{\bibinfo{volume}{121}}, \bibinfo{pages}{10,698}
  (\bibinfo{year}{2016}).

\bibitem[{\citenamefont{{Vernisse} et~al.}(2017)\citenamefont{{Vernisse},
  {Riousset}, {Motschmann}, and {Glassmeier}}}]{2017P&SS..137...40V}
\bibinfo{author}{\bibfnamefont{Y.}~\bibnamefont{{Vernisse}}},
  \bibinfo{author}{\bibfnamefont{J.~A.} \bibnamefont{{Riousset}}},
  \bibinfo{author}{\bibfnamefont{U.}~\bibnamefont{{Motschmann}}},
  \bibnamefont{and} \bibinfo{author}{\bibfnamefont{K.~H.}
  \bibnamefont{{Glassmeier}}}, \bibinfo{journal}{\planss}
  \textbf{\bibinfo{volume}{137}}, \bibinfo{pages}{40} (\bibinfo{year}{2017}).

\bibitem[{\citenamefont{{Neubauer}}(1999)}]{1999JGR...10428671N}
\bibinfo{author}{\bibfnamefont{F.~M.} \bibnamefont{{Neubauer}}},
  \bibinfo{journal}{\jgr} \textbf{\bibinfo{volume}{104}},
  \bibinfo{pages}{28671} (\bibinfo{year}{1999}).

\bibitem[{\citenamefont{{Kopp} and {Ip}}(2002)}]{2002JGRA..107.1490K}
\bibinfo{author}{\bibfnamefont{A.}~\bibnamefont{{Kopp}}} \bibnamefont{and}
  \bibinfo{author}{\bibfnamefont{W.-H.} \bibnamefont{{Ip}}},
  \bibinfo{journal}{Journal of Geophysical Research (Space Physics)}
  \textbf{\bibinfo{volume}{107}}, \bibinfo{eid}{1490} (\bibinfo{year}{2002}).

\bibitem[{\citenamefont{{Mottez} and {Zarka}}(2014)}]{2014A&A...569A..86M}
\bibinfo{author}{\bibfnamefont{F.}~\bibnamefont{{Mottez}}} \bibnamefont{and}
  \bibinfo{author}{\bibfnamefont{P.}~\bibnamefont{{Zarka}}},
  \bibinfo{journal}{\aap} \textbf{\bibinfo{volume}{569}}, \bibinfo{eid}{A86}
  (\bibinfo{year}{2014}), \eprint{1408.1333}.

\bibitem[{\citenamefont{{Caruso} et~al.}(2024)\citenamefont{{Caruso},
  {Buccino}, {Coffin}, {Gomez Casajus}, {Parisi}, {Zannoni}, {Gramigna},
  {Withers}, {Tortora}, {Park} et~al.}}]{2024EPSC...17..726C}
\bibinfo{author}{\bibfnamefont{A.}~\bibnamefont{{Caruso}}},
  \bibinfo{author}{\bibfnamefont{D.}~\bibnamefont{{Buccino}}},
  \bibinfo{author}{\bibfnamefont{D.}~\bibnamefont{{Coffin}}},
  \bibinfo{author}{\bibfnamefont{L.}~\bibnamefont{{Gomez Casajus}}},
  \bibinfo{author}{\bibfnamefont{M.}~\bibnamefont{{Parisi}}},
  \bibinfo{author}{\bibfnamefont{M.}~\bibnamefont{{Zannoni}}},
  \bibinfo{author}{\bibfnamefont{E.}~\bibnamefont{{Gramigna}}},
  \bibinfo{author}{\bibfnamefont{P.}~\bibnamefont{{Withers}}},
  \bibinfo{author}{\bibfnamefont{P.}~\bibnamefont{{Tortora}}},
  \bibinfo{author}{\bibfnamefont{R.~S.} \bibnamefont{{Park}}},
  \bibnamefont{et~al.}, in \emph{\bibinfo{booktitle}{European Planetary Science
  Congress}} (\bibinfo{year}{2024}), pp. \bibinfo{pages}{EPSC2024--726}.

\bibitem[{\citenamefont{{Smith} and {Goertz}}(1978)}]{1978JGR....83.2617S}
\bibinfo{author}{\bibfnamefont{R.~A.} \bibnamefont{{Smith}}} \bibnamefont{and}
  \bibinfo{author}{\bibfnamefont{C.~K.} \bibnamefont{{Goertz}}},
  \bibinfo{journal}{\jgr} \textbf{\bibinfo{volume}{83}}, \bibinfo{pages}{2617}
  (\bibinfo{year}{1978}).

\bibitem[{\citenamefont{{Mottez} and {Heyvaerts}}(2011)}]{2011A&A...532A..21M}
\bibinfo{author}{\bibfnamefont{F.}~\bibnamefont{{Mottez}}} \bibnamefont{and}
  \bibinfo{author}{\bibfnamefont{J.}~\bibnamefont{{Heyvaerts}}},
  \bibinfo{journal}{\aap} \textbf{\bibinfo{volume}{532}}, \bibinfo{eid}{A21}
  (\bibinfo{year}{2011}), \eprint{1106.0657}.

\bibitem[{\citenamefont{{Cairns}}(2004)}]{2004AIPC..719..381C}
\bibinfo{author}{\bibfnamefont{I.~H.} \bibnamefont{{Cairns}}}, in
  \emph{\bibinfo{booktitle}{Physics of the Outer Heliosphere}}, edited by
  \bibinfo{editor}{\bibfnamefont{V.}~\bibnamefont{{Florinski}}},
  \bibinfo{editor}{\bibfnamefont{N.~V.} \bibnamefont{{Pogorelov}}},
  \bibnamefont{and} \bibinfo{editor}{\bibfnamefont{G.~P.} \bibnamefont{{Zank}}}
  (\bibinfo{publisher}{AIP}, \bibinfo{year}{2004}), vol. \bibinfo{volume}{719}
  of \emph{\bibinfo{series}{American Institute of Physics Conference Series}},
  pp. \bibinfo{pages}{381--386}.

\bibitem[{\citenamefont{{Lyutikov}}(2006)}]{2006MNRAS.373...73L}
\bibinfo{author}{\bibfnamefont{M.}~\bibnamefont{{Lyutikov}}},
  \bibinfo{journal}{\mnras} \textbf{\bibinfo{volume}{373}}, \bibinfo{pages}{73}
  (\bibinfo{year}{2006}), \eprint{astro-ph/0604178}.

\bibitem[{\citenamefont{{Dursi} and {Pfrommer}}(2008)}]{2008ApJ...677..993D}
\bibinfo{author}{\bibfnamefont{L.~J.} \bibnamefont{{Dursi}}} \bibnamefont{and}
  \bibinfo{author}{\bibfnamefont{C.}~\bibnamefont{{Pfrommer}}},
  \bibinfo{journal}{\apj} \textbf{\bibinfo{volume}{677}},
  \bibinfo{eid}{993-1018} (\bibinfo{year}{2008}), \eprint{0711.0213}.

\bibitem[{\citenamefont{{Lyutikov}}(2023)}]{2023PhRvE.107b5205L}
\bibinfo{author}{\bibfnamefont{M.}~\bibnamefont{{Lyutikov}}},
  \bibinfo{journal}{\pre} \textbf{\bibinfo{volume}{107}}, \bibinfo{eid}{025205}
  (\bibinfo{year}{2023}), \eprint{2211.14433}.

\bibitem[{\citenamefont{{Sharma} et~al.}(2026)\citenamefont{{Sharma}, {Barkov},
  {Turyshev}, and {Lyutikov}}}]{Praveen}
\bibinfo{author}{\bibfnamefont{P.}~\bibnamefont{{Sharma}}},
  \bibinfo{author}{\bibfnamefont{M.~V.} \bibnamefont{{Barkov}}},
  \bibinfo{author}{\bibfnamefont{S.~G.} \bibnamefont{{Turyshev}}},
  \bibnamefont{and}
  \bibinfo{author}{\bibfnamefont{M.}~\bibnamefont{{Lyutikov}}}
  (\bibinfo{year}{2026}), \eprint{2602.14300}.

\bibitem[{\citenamefont{{Khlebnikov} and
  {Lyutikov}}(2025)}]{2025PhRvD.111j3010K}
\bibinfo{author}{\bibfnamefont{S.}~\bibnamefont{{Khlebnikov}}}
  \bibnamefont{and}
  \bibinfo{author}{\bibfnamefont{M.}~\bibnamefont{{Lyutikov}}},
  \bibinfo{journal}{\prd} \textbf{\bibinfo{volume}{111}}, \bibinfo{eid}{103010}
  (\bibinfo{year}{2025}).

\bibitem[{\citenamefont{{Lyutikov}}(2011)}]{2011PhRvD..83f4001L}
\bibinfo{author}{\bibfnamefont{M.}~\bibnamefont{{Lyutikov}}},
  \bibinfo{journal}{\prd} \textbf{\bibinfo{volume}{83}}, \bibinfo{eid}{064001}
  (\bibinfo{year}{2011}), \eprint{1101.0639}.

\bibitem[{\citenamefont{{Palenzuela} et~al.}(2010)\citenamefont{{Palenzuela},
  {Lehner}, and {Liebling}}}]{2010Sci...329..927P}
\bibinfo{author}{\bibfnamefont{C.}~\bibnamefont{{Palenzuela}}},
  \bibinfo{author}{\bibfnamefont{L.}~\bibnamefont{{Lehner}}}, \bibnamefont{and}
  \bibinfo{author}{\bibfnamefont{S.~L.} \bibnamefont{{Liebling}}},
  \bibinfo{journal}{Science} \textbf{\bibinfo{volume}{329}},
  \bibinfo{pages}{927} (\bibinfo{year}{2010}), \eprint{1005.1067}.

\bibitem[{\citenamefont{Arber et~al.}(2015)\citenamefont{Arber, Bennett, Brady,
  Lawrence-Douglas, Ramsay, Sircombe, Gillies, Evans, Schmitz, Bell
  et~al.}}]{Arber:2015hc}
\bibinfo{author}{\bibfnamefont{T.~D.} \bibnamefont{Arber}},
  \bibinfo{author}{\bibfnamefont{K.}~\bibnamefont{Bennett}},
  \bibinfo{author}{\bibfnamefont{C.~S.} \bibnamefont{Brady}},
  \bibinfo{author}{\bibfnamefont{A.}~\bibnamefont{Lawrence-Douglas}},
  \bibinfo{author}{\bibfnamefont{M.~G.} \bibnamefont{Ramsay}},
  \bibinfo{author}{\bibfnamefont{N.~J.} \bibnamefont{Sircombe}},
  \bibinfo{author}{\bibfnamefont{P.}~\bibnamefont{Gillies}},
  \bibinfo{author}{\bibfnamefont{R.~G.} \bibnamefont{Evans}},
  \bibinfo{author}{\bibfnamefont{H.}~\bibnamefont{Schmitz}},
  \bibinfo{author}{\bibfnamefont{A.~R.} \bibnamefont{Bell}},
  \bibnamefont{et~al.}, \bibinfo{journal}{Plasma Physics and Controlled Fusion}
  \textbf{\bibinfo{volume}{57}}, \bibinfo{pages}{1} (\bibinfo{year}{2015}).

\end{thebibliography}

\appendix

\section{Code  description, initial and boundary conditions}
\label{EPOCH}

We use user-modified PIC code EPOCH \citep{Arber:2015hc}.  For outer boundary conditions, we use {\it open}. 
Initial configuration involved a metal ball in external, non-penetrating  \Bf, parametrized by $B_0$, together with velocity flow corresponding to  incompressible flow around a sphere, parametrized by $v_0$. 
Initial \Ef\ is $\E_0 = - \v_0 \times \B_0$. At the  left  boundary $x_{\rm min}$  plasma is injected with $v_x =v_0$, $B_z =B_0$, $E_y =  v_0 B_0$. This set-up of the injection and initial conditions guarantees that  the injection is `smooth', as  at the sliding boundary between the initial and injected plasma the parameters (density, velocity and \EM\ fields) nearly match. 

We then introduce an internal conducting sphere, where we implement  perfectly conducting boundaries:
\begin{itemize} 
\item Tangential electric field is zero,  $E_{tan} = 0$, using  Applied field$\_$clamp$\_$zero.
\item Normal magnetic field is zero
\item Normal electric field: $E_{norm}^{in} = E_{norm}^{out}$,  allowing surface charge, using  field$\_$zero$\_$gradient
\item Tangential magnetic field: $B_{tan}^{in} = B_{tan}^{out}$,  allowing surface current
\end {itemize} 
(implementation of boundary conditions  turns out to be highly computationally intensive)

For example, for $v_0=0.5$, assumed density, and cold plasma,    the Alfvenic transition, $M_A=1$ occurs at  
$
B^{(M_A=1)}=0.5 \times 10^7\, {\rm  G} 
$.
For higher \Bfs\ the flow is relativistic but sub-Alfvenic.

 In  the code, for the basic run we set \Bf\ to 1 Tesla, size $1 \mu$m, density  $10^{24}$ $m^{-3}$) (EPOCH uses SI units).   For our basic run we use a box of $(128)^3$. Thus, both the skin depth and cyclotron radius are well resolved. 
  The \EM\ power falling on to the sphere is the $P =  2.4 \times 10^{15} $ erg s$^{-1}$, voltage difference of $3 \times 10^5 $ Volts ($10^3 $ statvolts.) For impedance of free space, the total current across the \NS\  is  $I_0=766 $ Amperes.  The current at $z\pm 3$ evaluates to $I_w = 183$ Amperes.

The code is run for  100 fs, $\sim 10$ dynamical time scale (of the whole box, not just the central sphere), to allow tor the system to settle into a stationary configuration. We verify 
field exclusion: $B$ field lines clearly drape around the sphere. High current density is  observed near the surface (diamagnetic currents).

\section{Road to parameters scan}

\begin{figure}[h!]
  \includegraphics[width=.49\linewidth]{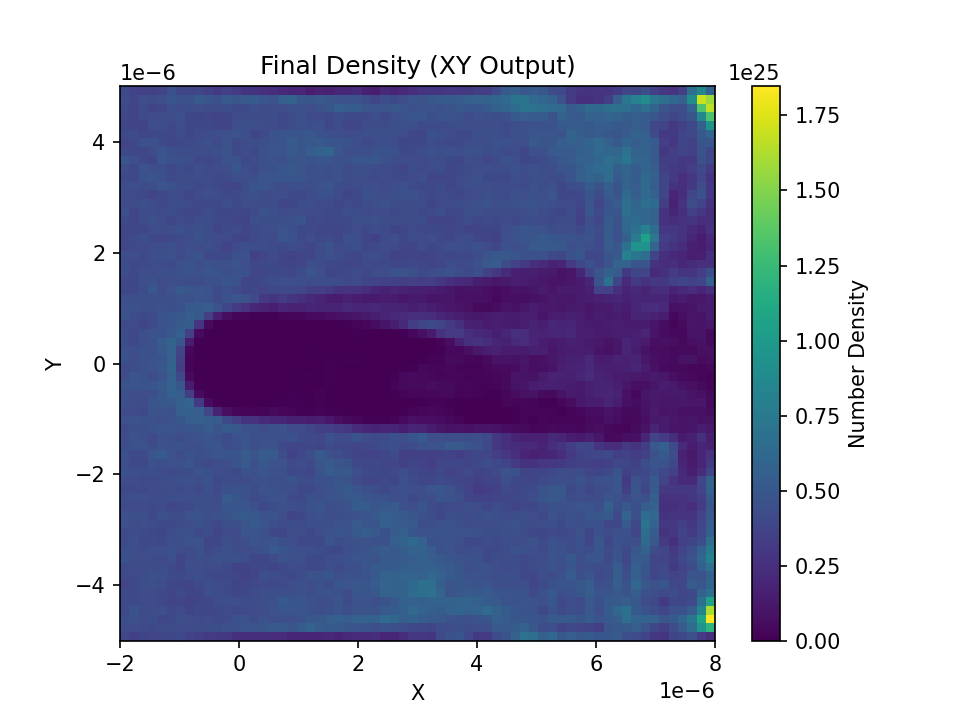} 
   \includegraphics[width=.49\linewidth]{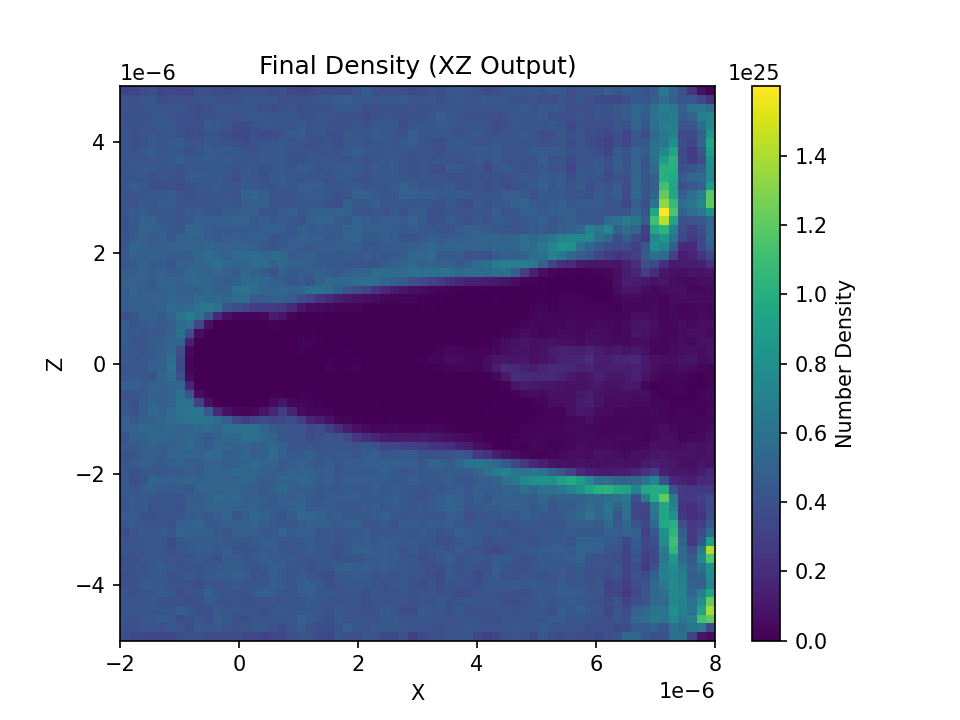} 
  \caption{Super-Alfvenic runs ($v_0=0.5,  \, B_0 =100 $ T, $M_A =2.6$). showing formation of a typical bow-shock structure. (Compare with Fig. \ref{slices1}). } 
 \label{lowB}
\end{figure}

Parameters' scan is complicated by a couple of issues, one physical one numerical.  The physical one is: for small \Bfs, the flow becomes super-Alfvenic, so instead of Alfven wings,  a bow shock structure develops, Fig. \ref{lowB}. 

Numerical issues are related to  the fact that for highly sub-Alfvenic runs the wings become wide-open, in all three directions. Wings' formation zone becomes large and starts to exceed the box size. This imposes heavy demands on the computing resources, as increasing box size while keeping resolution is expensive for 3D runs.  We postpone  the parameters scan for future work involving an order of magnitude larger numerical resources.

\section{Mimicking BH}
\label{MimickingBH}
Repeating the calculations for a BH moving across \Bf\ requires General Relativistic PIC code. We can still get a glimpse at the general properties by just changing the boundary conditions on the inner surface, from {\it  conducting}  to {\it open} (so that all the particles and fields inside the sphere are zeroed). This mimics the horizon of a BH as a spacial null surface.  The expectation  \citep{2011PhRvD..83f4001L} is that, surpassingly, the BH behaves similar to the NS.

\begin{figure}[h!]
  \includegraphics[width=.99\linewidth]{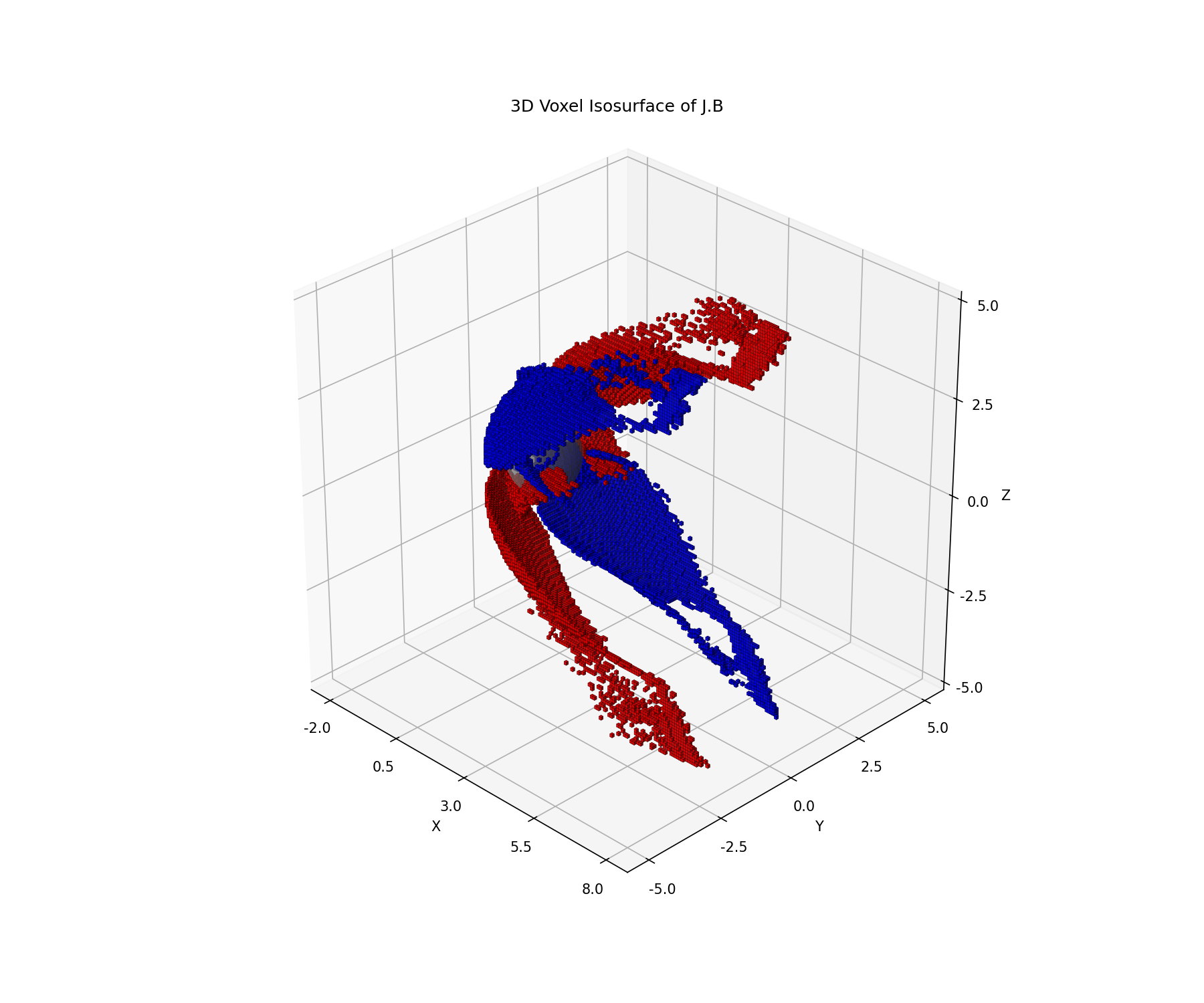}
   \caption{Mimicking the BH: 3D rendering of parallel current ({\it open} boundary condition on the inner surface). This set-up mimics the electromagnetic interaction of a \BH\ moving through magnetized plasma. Though the details differ from the conducting sphere case,  the general structure - formation of \Alfven wings  -  remains the same.}
    \label{BH}
\end{figure}

\begin{figure}[h!]
  \includegraphics[width=.35\linewidth]{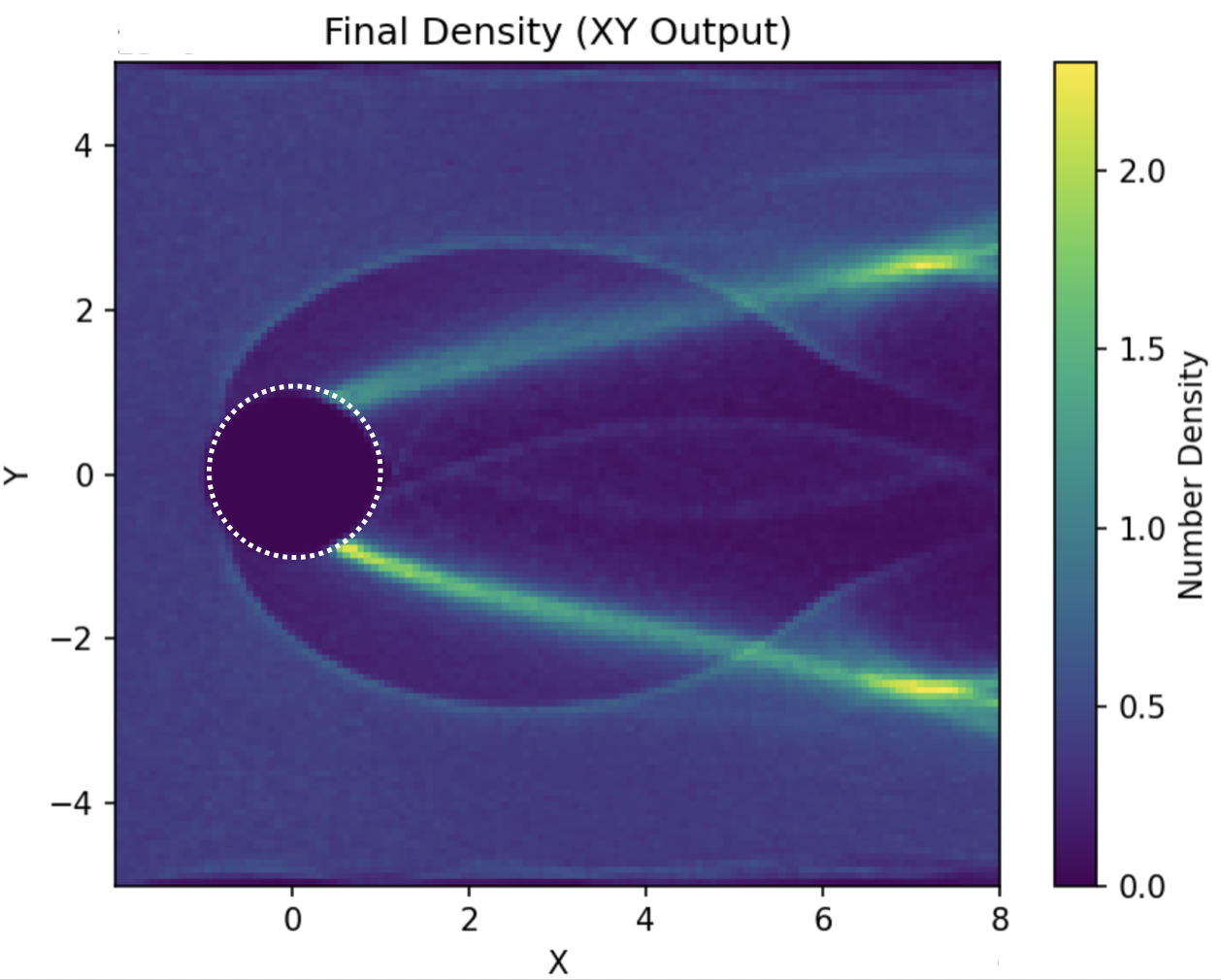} 
   \includegraphics[width=.35\linewidth]{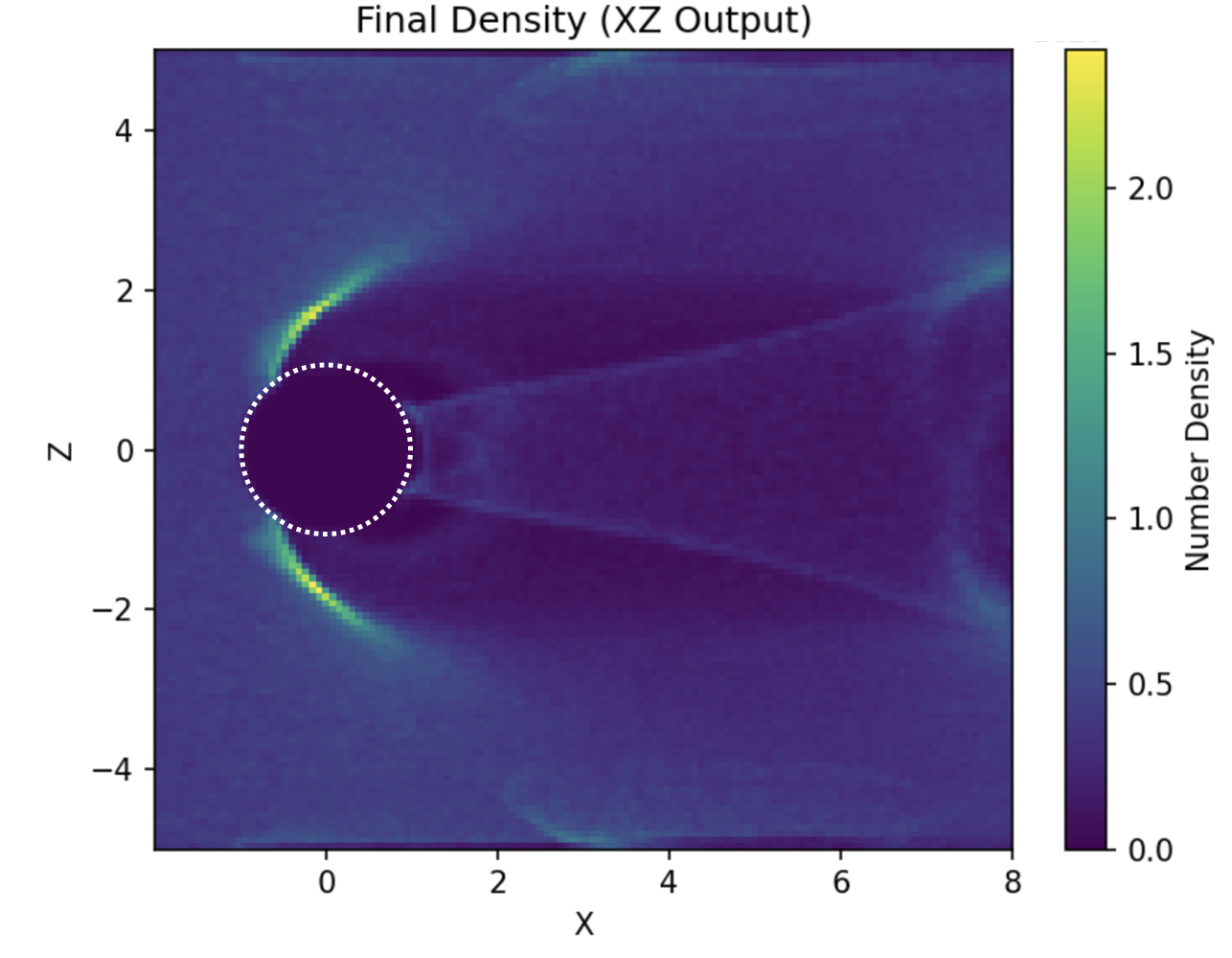} \\
     \includegraphics[width=.35\linewidth]{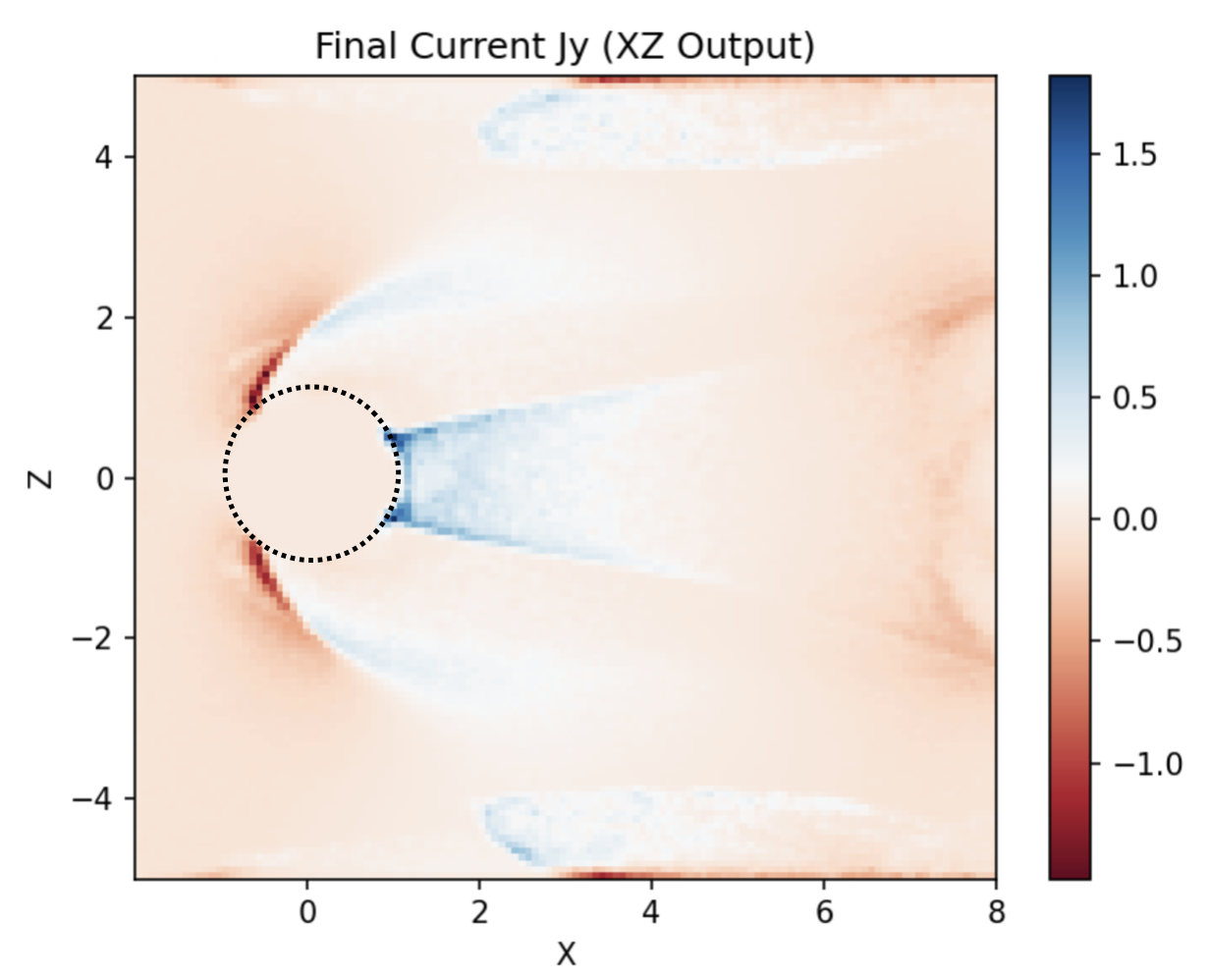} 
   \includegraphics[width=.35\linewidth]{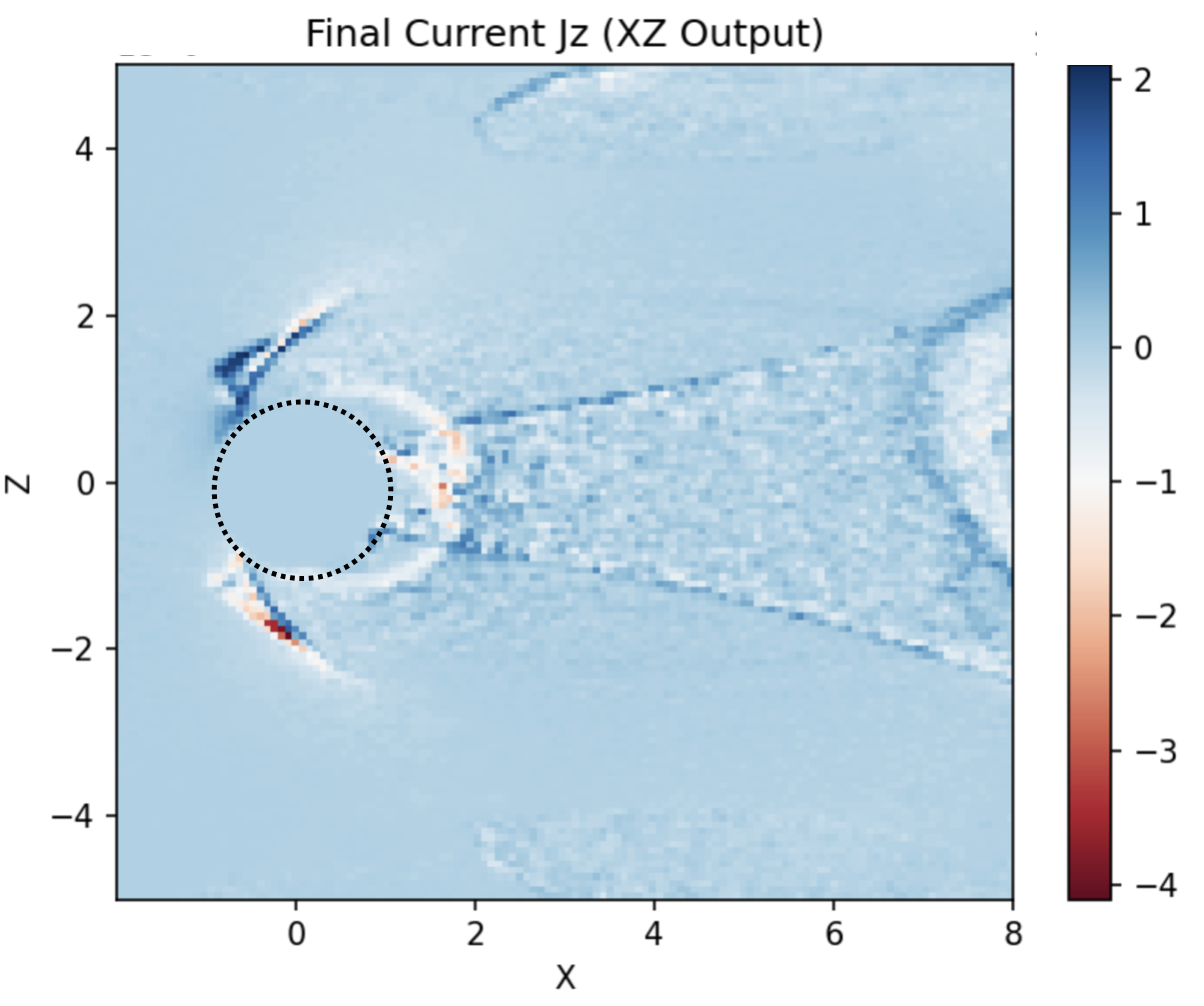} 
  \caption{Mimicking the BH:  density and current slices for {\it open} boundary condition on the inner surface.  } 
 \label{BH1}
\end{figure}

Our results conform with such expectation, Fig. \ref{BH}- \ref{BH1}. Though details differ, the overall structure - formation of  X-shaped current structure, \Alfven wings, remains.


 Beside being of independent physical interest, the case of mimicking the BH bears an important lesson:  the global properties  of  Alfven wings  in the relativistic  regime are generally independent  of the details of the boundary condition near the surface. This is  important: the code is likely under-resolves  the internal structure of the draping layer, which plays the role of the  source for the global currents. Yet, the structure of the global currents turn out to be only  weakly dependent on the detailed  structure of the source. One might invoke here a hydrodynamic analogy:  in the high Reynolds limit, the structure  of the turbulent wake, generated in the viscose Prandtl  boundary layer, turns out to be mostly independent of the detailed  structure of the viscose  sublayer.

\end{document}